\documentstyle[epsfig]{elsart}
\def\Journal#1#2#3#4{{#1} {\bf #2}, #3 (#4)}

\def\APP{{\em Acta Phys. Pol.} B}

\def\PLB{{\em Phys. Lett.}  B}
\def\PRL{\em Phys. Rev. Lett.}

\def\PRC{{\em Phys. Rev.} C}

\begin{document}
%\hspace{9.8 cm}NT@UW--99--??
\hfill{NT@UW-99-47}

\begin{frontmatter}
 
\title{Polarization phenomena in the reaction 
$NN \rightarrow NN\pi$ near threshold}
 
\author{
C. Hanhart$^a$, J. Haidenbauer$^b$, O. Krehl$^b$ and J. Speth$^b$}

\address{
\small{$^a$Dept. of Physics and INT, University of Washington,}\\
\small{Seattle, WA 98195-1560, USA} \\
\small{$^b$Institut f\"{u}r Kernphysik, Forschungszentrum J\"{u}lich
GmbH,}\\ 
\small{D--52425 J\"{u}lich, Germany} }
 
\maketitle

\begin{abstract}
First calculations for spin-dependent observables of the reactions 
$pp \rightarrow pp\pi^0$, $pp \rightarrow pn\pi^+$ and 
$pp \rightarrow d\pi^+$ near threshold
are presented, employing the J\"ulich model for pion production. The 
influence of resonant (via the excitation of the $\Delta (1232)$) and
non-resonant p-wave pion production mechanisms on these 
observables is examined. For the reactions $pp \rightarrow pn\pi^+$ and 
$pp \rightarrow d\pi^+$ nice agreement of our predictions with
the presently available data on spin correlation coefficents is
observed whereas for $pp \rightarrow pp\pi^0$ the description
of the data is less satisfying. 
\end{abstract}
\end{frontmatter}
 
\vskip 1cm 

The last few years witnessed a rather rapid growth of the data 
set on the various charge channels of the reaction 
$NN \rightarrow NN\pi$ near threshold \cite{DATA}. 
Naturally the first observables that became available were
total cross sections. But soon they were supplemented with
data on differential cross sections as well as analyzing
powers. At present a third stage has been reached where
results from measurements involving polarized beams as well as 
polarized targets are becoming available 
\cite{Mey98,Mey99,Sah99,Prz99,Jan00}.

As far as microscopic model calculations of the reaction $NN
\rightarrow NN\pi$ are concerned one has to concede that theory
is definitely lagging behind the developement on the experimental
sector. Many of the works \cite{DATA}
deal only with the reaction $pp \rightarrow pp\pi^0$. Furthermore
they take into account only the lowest partial wave(s).
Therefore it is not possible to confront those models with the
wealth of experimental information available nowadays, specifically
with differential cross sections and with spin-dependent observables.
In fact, to the best of our knowledge so far there are only two model 
calculations where all relevant pion production channels are considered and,
in addition, higher partial waves are included as well, namely the
ones of the Osaka \cite{Mae97,Tam98} and the J\"ulich \cite{Han98a,Han98b} 
groups.

The forthcoming data on spin-dependent cross sections and spin
correlation coefficients are very welcome since it is 
expected that they might play an important role in deepening
our theoretical understanding of pion production near threshold. 
Thus, in order to keep up with the developement on the experimental
side we want to present here corresponding predictions of the 
J\"ulich model in order to facilitate a comparison with the new 
measurements. Furthermore we investigate the
sensitivity of these observables to specific production mechanisms.
Such informations will be useful for a future more detailed analysis.
Note that this is the first time that model calculations of these 
observables are made available for the reactions 
$pp \rightarrow pp\pi^0$, $pp \rightarrow pn\pi^+$, and
$pp \rightarrow d\pi^+$ near threshold.

\begin{figure}[h]
\vspace{7cm}
\includegraphics{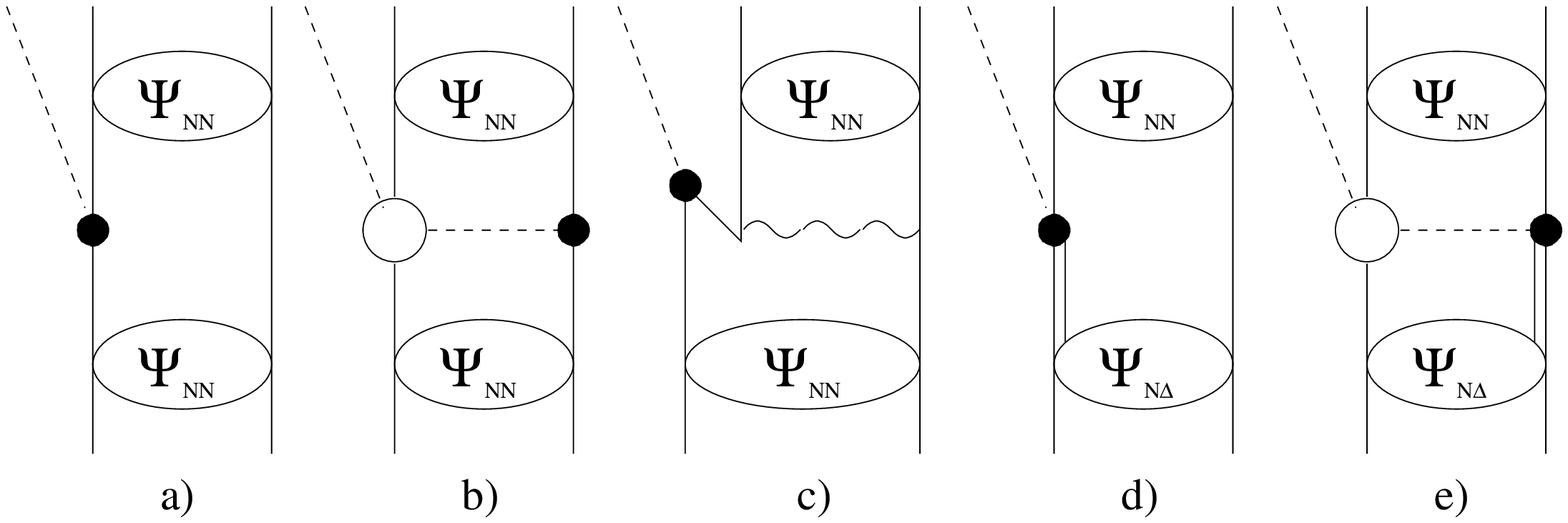}
\caption{\it Pion production mechanisms taken into account in our model:
(a) direct production; (b) pion rescattering; (c) contributions from pair
diagrams; (d) and (e) production involving the excitation of 
the $\Delta (1232)$ resonance. Note that diagrams where the 
$\Delta$ is excited after pion emission are also included.}
\label{beitraege} 
\end{figure}

Let us first describe shortly the J\"ulich model for pion production.
In this model all standard pion-production mechanisms
(direct production (Fig.~\ref{beitraege}a),
pion rescattering (Fig.~\ref{beitraege}b),
contributions from pair diagrams (Fig.~\ref{beitraege}c))
are considered. In addition, production mechanisms involving the
excitation of the $\Delta (1232)$ resonance 
(cf. Fig.~\ref{beitraege}d,e) are taken into account explicitly. 
All $NN$ partial waves up to orbital angular momenta $L_{NN} = 2$, and 
all states with relative orbital angular momentum $l \leq 2$ between 
the $NN$ system and the pion are considered in the final state. 
Furthermore all $\pi N$ partial waves up to orbital angular momenta
$L_{\pi N} = 1$ are included in calculating the rescattering diagrams
in Fig.~\ref{beitraege}b,e. Thus, our model includes not only
s-wave pion rescattering but also contributions from p-wave rescattering.

The reaction $NN \to  NN\pi$ is treated in a distorted wave born approximation,
in the standard fashion. The actual calculations are carried out 
in momentum space. For the distortions in the initial and final $NN$ states
we employ the model CCF of Ref.~\cite{HHJ}. This potential has been 
derived from the full Bonn model \cite{MHE87}
by means of the folded--diagram expansion. 
It is a coupled channel ($NN$, $N\Delta$, $\Delta\Delta$) model that
treats the nucleon and the $\Delta$ degrees of freedom on equal footing.
Thus, the $NN \leftrightarrow N\Delta$ transition amplitudes 
and the $NN$ T--matrices that enter in the evaluation of the pion
production diagrams in Fig. \ref{beitraege} are consistent solutions of 
the same (coupled--channel) Lippmann--Schwinger--like equation. 

The $\pi N \to \pi N$ T--matrix needed for the rescattering process is
taken from a microscopic meson--exchange model developed by the J\"ulich 
group \cite{SHH}. This interaction model is based on the conventional (direct 
and crossed) pole diagrams involving the nucleon and $\Delta$ isobar as 
well as t--channel meson exchanges in the scalar ($\sigma$) and vector
($\rho$) channel derived from correlated $2\pi$--exchange.  
Note that in our model of the reaction $NN\rightarrow NN\pi$
contributions where the pions are produced directly from the
nucleon or $\Delta$ (cf. Figs.\ref{beitraege}a and \ref{beitraege}d) 
are taken into account
explicitly. Therefore, the corresponding nucleon and $\Delta$ pole terms 
have to be taken out of the $\pi N$ T--matrix in order to avoid double 
counting. 

The contributions of the pair diagrams (Fig.~\ref{beitraege}c) are 
viewed as an effective parametrization of short range production
mechanisms that are not explicitly included in the model. 
Their strength, the only free parameter in the J\"ulich model,
was adjusted to reproduce the total $pp\pi^0$ production 
cross section at low energies. Note that, due to their vertex
structure, those pair diagrams contribute only to s-wave pion
production. 
 
Results of this model for total cross sections and analyzing powers
for the reactions channels
$pp \rightarrow pp\pi^0$, $pp \rightarrow pn\pi^+$,
$pn \rightarrow pp\pi^-$, and $pp \rightarrow d\pi^+$ were
presented in Refs.~\cite{Han98a,Han98b}. It was found that the model 
yields a very good overall description of the data from the 
threshold up to the $\Delta$ resonance region. 
In fact, a nice quantitative agreement with basically all experimental
information (then available) was observed over a wide energy range.
Thus, this model is very well suited as a starting point for a detailed
analysis of the forthcoming spin-dependent observables of the reaction 
$NN \rightarrow NN\pi$.

\begin{figure}[h]
\vspace{13cm}
\includegraphics{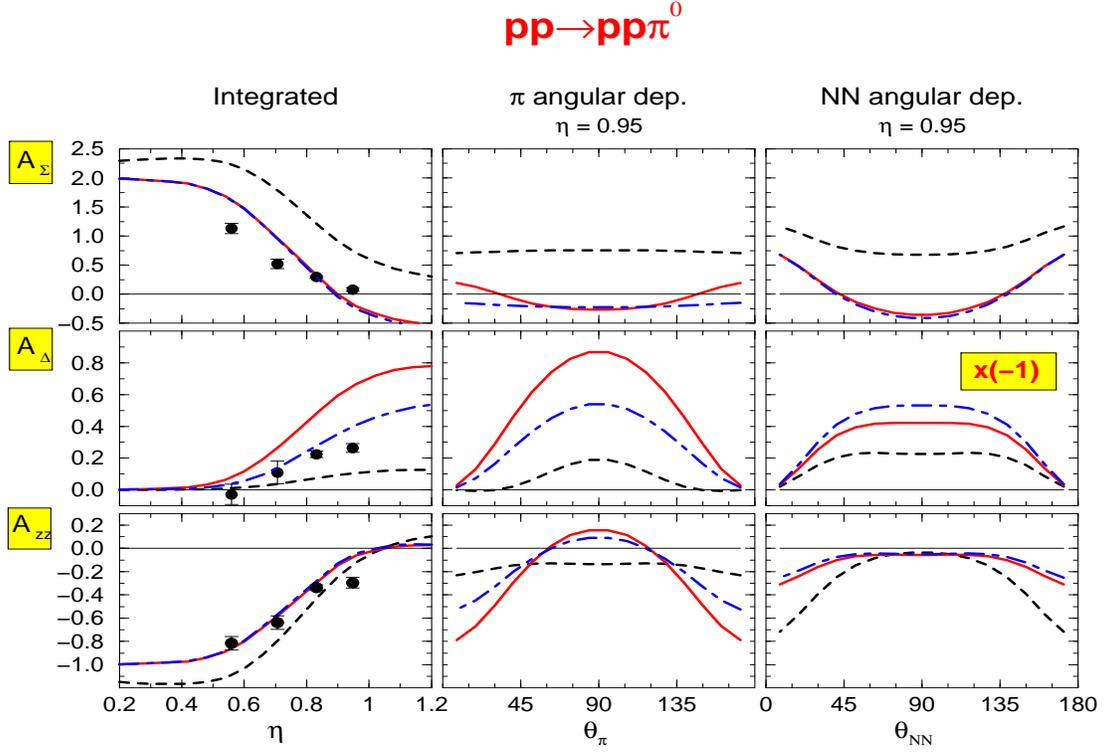}
\caption{\it {Spin correlation paramaters for the reaction
$pp \to pp\pi^0$, where $A_\Sigma = A_{xx}+A_{yy}$ and 
$A_\Delta = A_{xx}-A_{yy}$. The solid line represents the result
of our full model. The dashed-dotted curves are obtained when the
contributions of non-resonant p-wave pion rescattering are omitted.
The dashed curves show results where the contributions to
pion production via the $\Delta (1232)$ resonance are switched off
as well. The experimental results are taken from Ref.~\protect\cite{Mey98}. 
}}
\label{Corr1}
\end{figure}

\begin{figure}[h]
\vspace{13cm}
\includegraphics{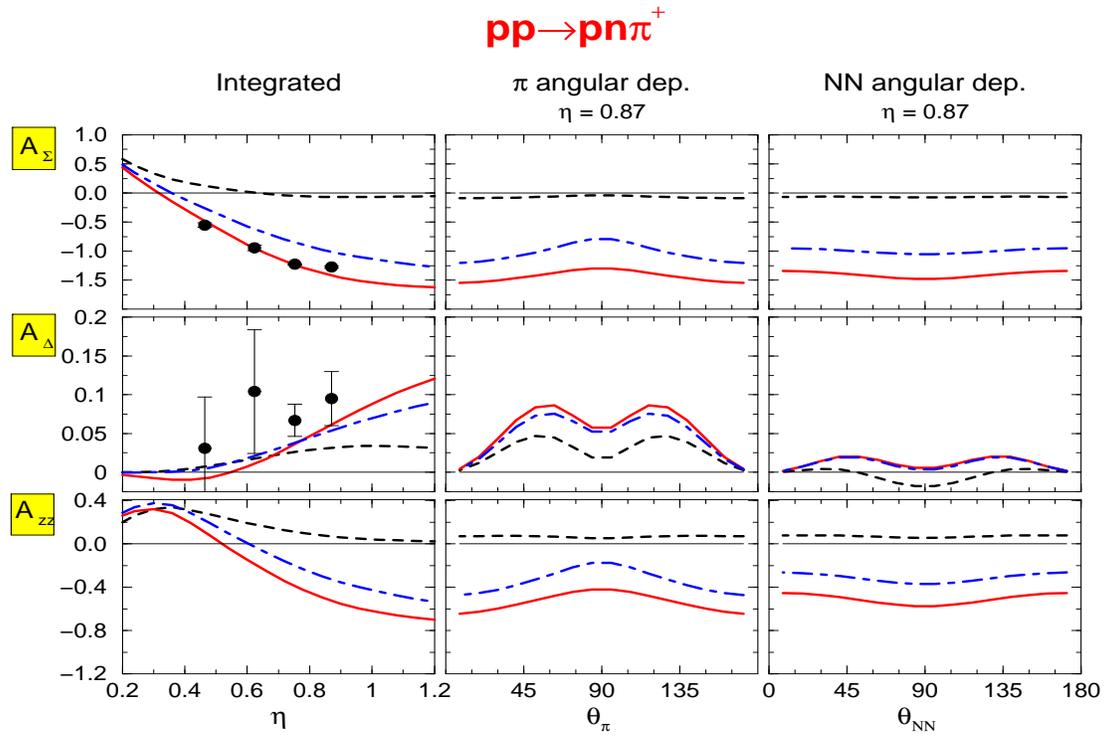}
\caption{\it{Spin correlation paramaters for the reaction
for $pp \to pn\pi^+$. Same description of the curves as in 
Fig.~\ref{Corr1}.  The experimental results are taken from 
Ref.~\protect\cite{Sah99}. }} 
\label{Corr2}
\end{figure}

Predictions for the spin correlation coefficient combinations 
$A_\Sigma = A_{xx}+A_{yy}$, $A_\Delta = A_{xx}-A_{yy}$ and $A_{zz}$ 
are shown in Figs.~\ref{Corr1} (for $pp\rightarrow pp\pi^0$), 
\ref{Corr2} (for $pp\rightarrow pn\pi^+$), and 
\ref{Corr3} (for $pp\rightarrow d\pi^+$). The polar integrals of 
these observables are displayed in the left panels as a function of 
$\eta$, the maximum momentum of the produced pion in units of the pion mass.
(Note that the polar integral of $-(A_{xx}+A_{yy})$ and $A_{zz}$ yield the
spin-dependent total cross section $\Delta \sigma_T/\sigma_{tot}$ 
and $\Delta \sigma_L/\sigma_{tot}$, respectively; 
cf. Refs.~\cite{Mey98,Meyba} for definitions.) 
The other two panels contain the results 
at $T_{lab}$ = 400 MeV as a function of the pion angle (middle
panel) and of the angle between the nucleons (right panel). 

One of the specific features of the J\"ulich model is that 
contributions from p-wave pion rescattering are fully taken into 
account. Their resonant part is, of course, given by the 
pion production via the $\Delta$ excitation as depicted in 
Fig.~\ref{beitraege}d. However our model
includes non-resonant contributions from p-wave pion rescattering
as well. Thus, we can study the influence
of the latter on those spin observables. Results were the
contributions of non-resonant p-wave pion rescattering are omitted
are shown by the dash-dotted curves in Figs.~\ref{Corr1}-\ref{Corr3}.
One can see that the effect of these contributions is 
definitely not negligible. E.g., there is a strong influence on 
the observable $A_\Delta$, which is visible in the angular dependence 
as well as in the integrated result. In the reactions
$pp\rightarrow pn\pi^+$ and $pp\rightarrow d\pi^+$
non-resonant p-wave rescattering even yields a change in the sign
for energies $\eta \leq 0.6$. But since in this energy range the
overall magnitude of $A_\Delta$ is rather small it will be difficult 
to resolve these differences experimentally. 
Also the other spin correlation coefficient combinations are, 
in general, significantly modified by the contributions from 
non-resonant p-wave rescattering, in particular at higher energies. 

The dashed curves in Figs.~\ref{Corr1}-\ref{Corr3} 
represent results where the contributions from pion production via 
$\Delta$ excitation are switched off as well. 
Evidently this leads to rather large changes manifesting
the important role which the $\Delta$ plays for these spin-correlation
parameters. Thus, these observables are very well suited for testing
the model treatment of the pion-production contributions involving
the $\Delta$ resonance. In particular, they allow to examine the 
$\Delta$ contributions at energies far below the resonance regime. 
As can be seen from Figs.~\ref{Corr1}-\ref{Corr3}, the effect of the 
$\Delta$ extends down to fairly low energies, specifically in the 
reaction $pp\rightarrow pp\pi^0$.

Note that all results shown in Figs.~\ref{Corr1}-\ref{Corr3} are 
normalized to the same total cross section (for each reaction), 
namely the one predicted by the full model.
Without this re-normalization dramatic but basically artificial changes 
in the spin correlation coefficients would appear when adding additional 
contributions. 

For all three considered reactions some experimental information
on spin correlation coefficients has become available 
very recently and thus we can already compare the predictions 
of our model with them. 
In case of the reaction $pp\rightarrow pp\pi^0$ there will be
data soon on all the observables shown in Fig.~\ref{Corr1} \cite{Jan00}. 
So far values for
the integrated spin-correlation coefficients $A_\Sigma$, $A_\Delta$ 
and $A_{zz}$ have been published \cite{Mey98}.
Evidently two of those observables are reasonably well reproduced by
our model calculation (cf. Fig.~\ref{Corr1}) whereas 
$A_\Delta$ is overestimated by a factor 2 or so. Since the result without
non-resonant p-wave $\pi N$ rescattering (dash-dotted curve) goes almost 
through the data points one might be inclined to conclude that their
contributions are much too large in our model, as we argued in 
Ref.~\cite{Han98a}. However, one has to keep in mind that for energies
corresponding to $\eta \approx 1$ our model underestimates the total 
$\pi^0$ production cross section already by a factor of 2 or so 
(cf. Fig. 3 in Ref.~\cite{Han98a}). 
Since the spin correlation coefficients are normalized by $\sigma_{tot}$
it's conceivable that the disagreement with the data for $A_\Delta$ 
simply reflects the shortcoming in the total cross section. 
In order to understand this let us remind the reader that the 
numerator of $A_\Delta$ is basically determined by the $Pp$ partial 
waves \cite{Meyba}. 
(We use here the standard nomenclature for labelling the 
amplitudes by the angular orbital momentum in the final 
$NN$ system and of the pion relative to the $NN$ system.)
The deficiency in the total cross section, on the other hand, 
could be due to the $Ss$ amplitude, which might be
too small at larger energies in our model. Thus, 
an enhancement in the $Ss$ amplitude would lead to an increase in the
total cross section (as required by the data) and accordingly to
an increase in the denominator of $A_\Delta$. But it would not
change the numerator of $A_\Delta$ so that one would get an
overall reduction of $A_\Delta$. 
 
Note that in case of the other spin correlation coefficents
the $Ss$ amplitude enters in the denominator as well as in 
the numerator so that they should be less affected by the 
aforementioned deficiency in the total cross section. 

For the reaction $pp\rightarrow pn\pi^+$ there are data 
on the integrated spin-correlation coefficients $A_\Sigma$ and
$A_\Delta$ \cite{Sah99}. The former observable is very nicely
described by our model prediction (Fig.~\ref{Corr2}). It is
interesting that contributions from pion-production via the
$\Delta$ resonance as well as from (non-resonant) p-wave
rescattering are obviously required for achieving this agreement.
With the $\Delta$ resonance alone the result would lie clearly
below the experiment (cf. the dash-dotted curve). Our model is
also in rough agreement with the data on $A_\Delta$. Here, however,
the large error bars do not really allow to draw more quantitative
conclusions. 

Finally, for $pp\rightarrow d\pi^+$ there are angular distributions
for the spin-correlation coefficients $A_\Sigma$, $A_\Delta$, 
and $A_{zz}$ \cite{Prz99}. Also these data are nicely reproduced
by our model, cf. Fig.~\ref{Corr3}. Note that again the contributions 
from (non-resonant) p-wave rescattering are crucial for getting 
agreement with the data on $A_\Sigma$. 

In summary, we have presented first calculations of spin correlation 
coefficients 
for the reactions $pp \rightarrow pp\pi^0$, $pp \rightarrow pn\pi^+$,
and $pp \rightarrow d\pi^+$ near threshold. 
We have also studied the influence of resonant (i.e. via the 
$\Delta (1232)$ excitation) and non-resonant p-wave pion production 
mechanisms on these observables. 
Our model calculation is in rather good agreement with the presently
available data for the reactions $pp \rightarrow pn\pi^+$
and $pp \rightarrow d\pi^+$. This is certainly remarkable. We want
to emphasize again that our results are genuine model predictions. 
For the reaction $pp \rightarrow pp\pi^0$, however, the description
of the data is less satisfying. In particular, for the
spin correlation coefficent combination $A_{xx}-A_{yy}$ there is
even a serious disagreement with the experimental evidence. 
Here further investigations are required. Specifically it will be
interesting to see whether this deficiency is connected with the
still unsettled issue of the missing (s-wave) strength in the 
$pp\rightarrow pp\pi^0$ total cross section \cite{DATA} or
whether it is a sign for an additional problem concerning now the 
p-wave contributions to this reaction channel.

\vskip 1cm 

{\bf Acknowledgments}

C.H. is grateful for financial support by the Department of Energy
grant DE-FG03-97ER41014 and by the Alexander-von-Humboldt 
Foundation

\vfill \eject

\vfill \eject

\vglue 1cm  

\vfill \eject

\vglue 1cm  

\vfill \eject

\vglue 1cm  
\begin{figure}[h] 
\vspace{13cm}
\includegraphics{pptodpipl_pol.epsi}
\caption{\it{Spin correlation paramaters for the reaction
for $pp \to d\pi^+$. Same description of the curves as in 
Fig.~\ref{Corr1}.  The experimental results are taken from 
Ref.~\protect\cite{Prz99}. }} 
\label{Corr3}
\end{figure}

\end{document}